\documentclass[aps, pre, preprint]{revtex4-2}
\usepackage[english]{babel}
\usepackage{bm}
\usepackage{amsmath}
\usepackage{natbib}
\usepackage{graphicx}

\begin{document}

\title{Twist-induced local curvature of filaments in DNA toroids}

\author{Luca Barberi}
\altaffiliation[Present address: ]{Department of Biochemistry, University of Geneva, 1211 Geneva, Switzerland}
\affiliation{Université Paris-Saclay, CNRS, LPTMS, 91405, Orsay, France}

\author{Martin Lenz}
\affiliation{Université Paris-Saclay, CNRS, LPTMS, 91405, Orsay, France}
\affiliation{PMMH, CNRS, ESPCI Paris, PSL University, Sorbonne Université, Université de Paris, F-75005, Paris, France}

\begin{abstract}
	DNA toroidal bundles form upon condensation of one or multiple DNA filaments. DNA filaments in toroidal bundles are hexagonally packed, and collectively twist around the center line of the toroid. In a previous study, we and our coworkers argue that the filaments' curvature locally correlates with their density in the bundle, with the filaments less closely packed where their curvature appears to be higher. We base our claim on the assumption that twist has a negligible effect on the local curvature of filaments in DNA toroids. However, this remains to be proven. We fill this gap here, by calculating the distribution of filaments' curvature in a geometric model of twisted toroidal bundle, which we use to describe DNA toroids by an appropriate choice of parameters. This allows us to substantiate our previous study and suggest directions for future experiments.
\end{abstract}

\maketitle

\section{Introduction}

DNA is a negatively charged polymer. While its subunits repel each other in neutral solutions, they can effectively attract each other when in the presence of positive ions with high valency, thanks to a phenomenon known as DNA condensation \cite{grosberg2002}. When one or multiple DNA filaments condense, they most commonly aggregate into toroidal bundles. In DNA toroids, filaments are laterally packed as in a hexagonal lattice, and the whole lattice rotates around the toroid center line \cite{leforestier2009}. We provide a pictorial representation of such bundle in Fig.~\ref{fig:figure1}.

In a recent article, we and our coworkers use cryoelectron microscopy to reveal a novel structural feature of DNA toroids. Along the direction denoted by $\bm{e}_R$ in Fig.~\ref{fig:figure1}, the distance between neighboring filaments decreases with their distance from the $z$-axis \cite{barberi2020}. In the same article, we explain this behavior by means of a simple theory of mechanical equilibrium, based on the competition between local filament-filament interactions and DNA bending rigidity. The key mechanism in our theory is that the elastic energy penalty due to DNA bending rigidity promotes the accumulation of filaments in the regions of the bundle where their curvature is low. We choose to neglect the collective twist of DNA filaments around the toroid center line, \emph{i.e.} to describe them as circles lying perpendicular to the $z$-axis. In this case, curvature clearly decreases with the distance from the $z$-axis, producing a distribution of filaments consistent with the experiments at mechanical equilibrium.

Our prediction in Ref.~\cite{barberi2020} relies on the assumed distribution of filaments' curvature in the bundle, which neglects twist. However, twist can spatially redistribute curvature within the bundle, possibly invalidating our assumption. This is the problem we tackle here. In Sec.~\ref{sec:twist}, we introduce a simple geometric model of twisted toroidal bundle, which allows us to determine the expected distribution of the filaments' curvature as a function of twist. Then, in Sec.~\ref{sec:discussion}, we adapt our model to the case of DNA toroids and assess the validity of the assumption used in Ref.~\cite{barberi2020}.

\section{Geometric model of twisted toroidal bundle}\label{sec:twist}

\begin{figure}
\centering
\includegraphics[width = 0.5\textwidth]{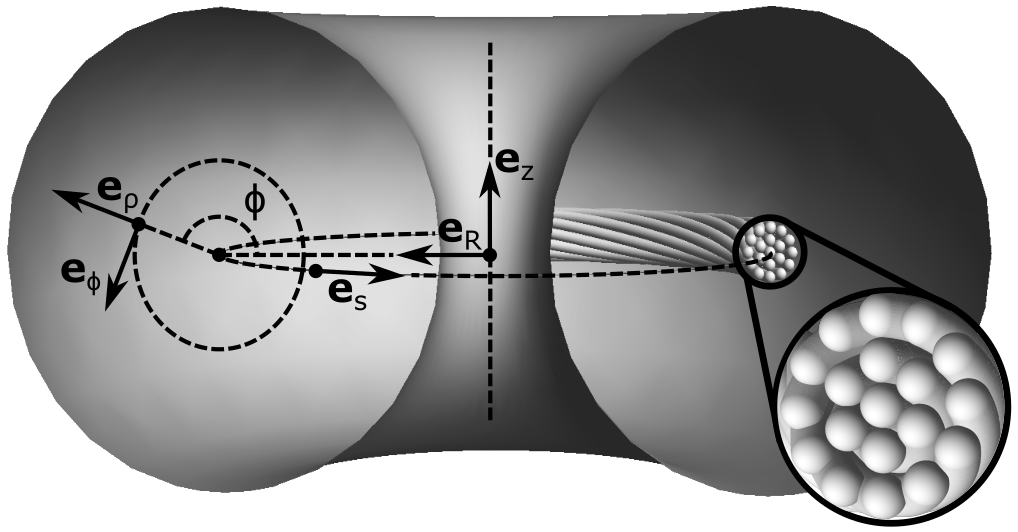}
\caption{Geometry of a twisted toroidal bundle. On the left, the reference frame used to locate points within a toroidal volume. On the right, a pictorial representation of a discrete bundle of filaments, which features the lateral hexagonal packing observed in DNA toroids.}\label{fig:figure1}
\end{figure}

We model a toroidal bundle as a continuous family of curves that surrounds a circular center line with radius $R$, following the approach used in Ref.~\cite{atkinson2019}. The center line is a curve with arc length $s \in [0, 2\pi R]$ and position vector $\bm{R}(s) = R \,\bm{e}_R(s)$ (Fig.~\ref{fig:figure1}). To describe the space surrounding the center line, we introduce the local frame of reference $(\bm{e}_s, \bm{e}_\rho, \bm{e}_\phi)$, where $\bm{e}_s = \partial_s \bm{R}$, $\bm{e}_\rho = -\cos\phi \,\bm{e}_R + \sin\phi \,\bm{e}_z$, $\bm{e}_\phi = \partial_\phi \bm{e}_\rho$ (Fig.~\ref{fig:figure1}). This allows us to locate each point $\bm{x}$ in space in terms of its distance $\rho$ from the center line, $\bm{x}(s, \rho, \phi) = \bm{R}(s) + \rho \,\bm{e}_\rho(s, \phi)$.

The geometry of the bundle is fully determined by the field $\bm{e}_t(s, \phi, \rho)$, which locally corresponds to the unit tangent of the filament passing through $\bm{x}(s, \phi, \rho)$. We prescribe a form of $\bm{e}_t$ that implicitly accounts for the chiral interactions between neighboring DNA filaments. We take inspiration from the theory of cholesteric liquid crystals, which predicts that dense packings of screw-like objects self-organize in a so-called \emph{double-twist} geometry \cite{wright1989}. In a straight bundle featuring double-twist, filaments are helices with pitch $2\pi/\Omega_s$ and tilt angle $\theta = \arctan(\Omega_s\rho)$ with respect to the central axis of the bundle. To mimic the double-twist geometry in our toroidal bundle, we define the field of tangents as
\begin{align}\label{eq:doubletwist}
	\bm{e}_t = \cos[\theta(\rho, \phi)] \,\bm{e}_s + \sin[\theta(\rho, \phi)] \,\bm{e}_\phi .
\end{align}
Assuming rotational invariance around the $z$-axis implies that $\theta(\rho, \phi)$ does not depend on $s$. Further assuming that DNA bundles are incompressible implies the no-splay condition $\nabla \cdot \bm{e}_t = 0$, where $\nabla$ is the gradient operator, which is satisfied by \cite{kulic2004, barberi2019}
\begin{align}\label{eq:splayfree}
	\theta(\rho, \phi) = \arcsin\left(\frac{f(\rho)}{1 - \rho\cos(\phi)/R}\right) ,
\end{align}
where $f(\rho)$ is an arbitrary function of $\rho$. In the following, we make the simplest linear choice $f(\rho) = \Omega\rho$, consistent with previous works on twisted toroids (for instance, see Ref.~\cite{kulic2004}). Note that the limited domain of the arcsin in Eq.~\eqref{eq:splayfree} imposes an upper bound on the tubular radius, $\rho < \rho^\text{max}_\Omega = R/(1 + \Omega R)$.

The parameter $\Omega$ governs the amount of twist in our bundle. It is related to the number of turns a filament makes around the toroid center line, but unfortunately not through a simple relation \cite{atkinson2019}. The case $\Omega = 0$ corresponds to an untwisted bundle, and the sign of $\Omega$ determines the handedness of the twist.

\begin{figure}
	\centering
	\includegraphics[width = \textwidth]{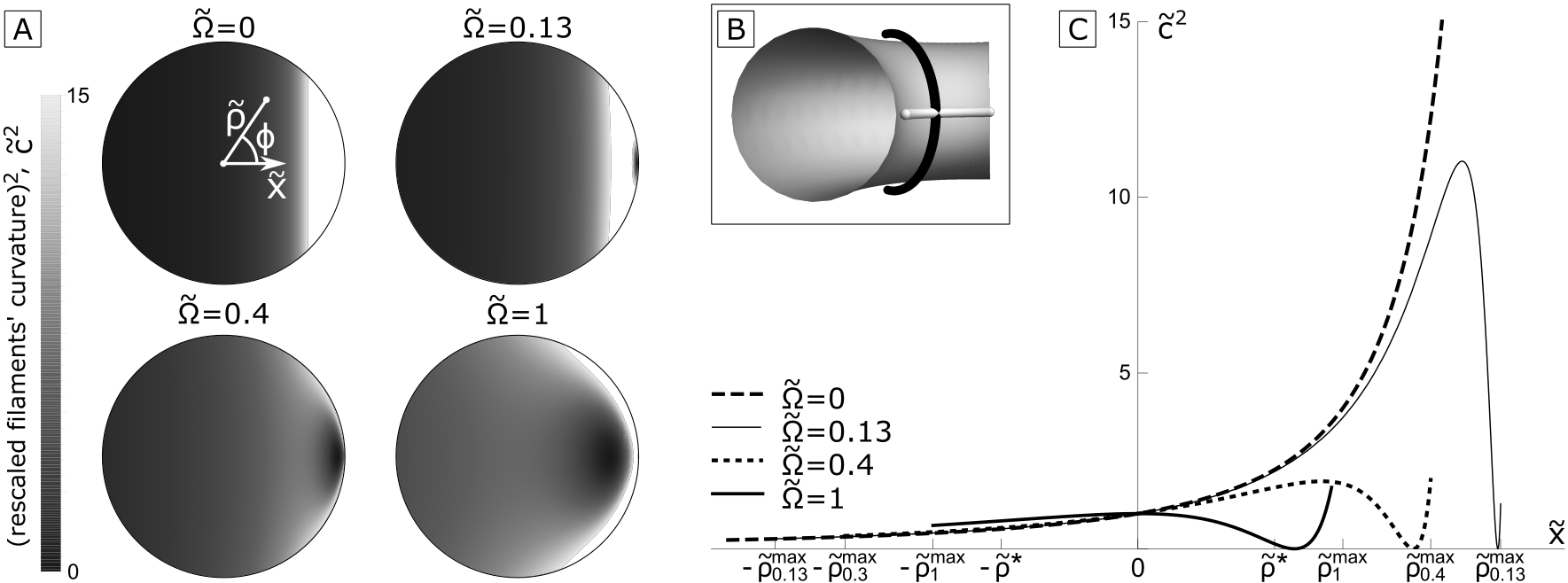}
	\caption{Curvature distribution in a twisted toroid. (A) Distribution of filaments' curvature at different values of $\widetilde{\Omega}$, over a two-dimensional domain $\phi \in [0, 2\pi] \times \widetilde{\rho} \in [0, \widetilde{\rho}^\text{max}_{\widetilde{\Omega}}]$. To facilitate view, cross-sections are rescaled so as to be equally sized. (B) Two arcs with curvature of opposite sign, on a torus. (C) Distribution of filaments' curvature at different values of $\widetilde{\Omega}$, over the $x$-axis defined in panel A. The domain of each curve is $[-\widetilde{\rho}^\text{max}_{\widetilde{\Omega}},\widetilde{\rho}^\text{max}_{\widetilde{\Omega}}]$, which depends on $\widetilde{\Omega}$.}\label{fig:figure2}
\end{figure}

From now on, we use $R$ as unit of length and denote nondimensional quantities by a tilde. Combining Eqs.~\eqref{eq:doubletwist} and \eqref{eq:splayfree} determines the form of $\bm{e}_t$, which yields the field of filaments' local curvature $\widetilde{c}(\widetilde{\rho}, \phi)$ through the relation $\widetilde{c}\,\bm{e}_n = (\bm{e}_t \cdot \nabla) \,\bm{e}_t$, where $\bm{e}_n$ is the field of local unit normals to the filaments. Rather than detailing its cumbersome expression \cite{barberi2019}, in Fig.~\ref{fig:figure2}A we plot $\widetilde{c}^2$ over the cross-section of a toroid with tubular radius $\widetilde{\rho}^\text{max}_{\widetilde{\Omega}}$, for different values of $\widetilde{\Omega}$. Note that the profile of $\widetilde{c}^2$ over a cross-section does not depend on the tubular radius.

\section{Conclusion}\label{sec:discussion}
At $\widetilde{\Omega} = 0$, the bundle does not twist and filaments are circles perpendicular to the $z$-axis. In this case, curvature decreases with the distance from the $z$-axis (Fig.~\ref{fig:figure2}A), as in our previous paper \cite{barberi2020}. If $\widetilde{\Omega} \neq 0$, twist alters the spatial distribution of curvature. At $\widetilde{\Omega} > 0$, a local curvature drop emerges at the boundary of the bundle cross section, close to the $z$-axis (Fig.~\ref{fig:figure2}A). This low-curvature region forms due to a local change of sign of the filaments' curvature. In fact, as $\widetilde{\Omega}$ increases, filaments transition from a state of ``positive'' curvature, in which they wind around the $z$-axis like the white arc in Fig.~\ref{fig:figure2}B, to a state of ``negative'' curvature, in which they wind around the toroid center line like the black arc in Fig.~\ref{fig:figure2}B, passing through a state of zero curvature at the transition. As $\widetilde{\Omega}$ increases, this local curvature drop shifts along the $x$-axis (Fig.~\ref{fig:figure2}A, C), towards the center of the bundle cross section.

We now estimate the distribution of curvature in the DNA toroids of Ref.~\cite{leforestier2009}. These toroids have a radius $R^* \simeq 25\,\text{nm}$ and (nondimensional) tubular radius $\widetilde{\rho}^* = \rho^*/R^* \simeq 1/3$. Differently than our model, the rate of twist of DNA bundles changes along the toroid center line, most of the twist being localized in so-called \emph{twist walls} \cite{leforestier2009}. In between twist walls, the bundle rotates rather slowly, and we showed elsewhere that the experimental data is consistent with a local $\widetilde{\Omega}^* = \Omega^* R^* \simeq 0.13$ \cite{barberi2019}. Assuming this amount of twist, our analysis suggests that the distribution of curvature within a toroid of tubular radius $\widetilde{\rho}^*$ is analogous to that of the $\widetilde{\Omega} = 0$ case (Fig.~\ref{fig:figure2}C). Note that $\widetilde{\rho}^* < \widetilde{\rho}^\text{max}_{\widetilde{\Omega} = 0.13}$, therefore the regions with sizeable deviations between the values of $\widetilde{c}^2$ of the $\widetilde{\Omega}=0$ and $\widetilde{\Omega}=0.13$ models in Fig.~\ref{fig:figure2}C are not a concern for us.

To conclude, our results substantiate the no-twist approximation used in Ref.~\cite{barberi2020} that the curvature of DNA filaments decreases with the distance from the center of the toroid. This hypothesis may fail close to twist walls, where the increased local twist may induce a non-monotonic dependence of the filaments' curvature on the distance from the center. If curvature and spacing correlate, as proposed in Ref.~\cite{barberi2020}, close to twist walls we may expect a non-monotonic dependence of the filaments spacing on the distance from the center. In principle, this possibility could be investigated in future experiments, similar to those in Ref.~\cite{barberi2020}. 

\section*{Acknowledgments}
LB thanks Gregory M. Grason and Daria W. Atkinson for helping in the development of the geometric model. LB was supported by the “IDI 2016” project funded by the IDEX Paris-Saclay, ANR-11-IDEX-0003-02. ML was supported by Marie Curie Integration Grant PCIG12-GA-2012-334053, “Investissements d’Avenir” LabEx PALM (ANR-10-LABX-0039-PALM), ANR grant ANR-15-CE13-0004-03 and ERC Starting Grant 677532. ML’s group belongs to the CNRS consortium CellTiss.

\end{document}